\documentclass[12pt,a4paper]{article}

 \usepackage[dvips]{graphics}
 \usepackage{graphicx}
 \usepackage{afterpage}
 \usepackage{enumerate}
  \usepackage{longtable}
\begin{document}
 %%==========================================================

 \title{\bf Parameters of sensitivity of Fraunhofer lines \\
 to changes in the temperature, gas pressure, \\
 and microturbulent velocity \\
 in the solar photosphere }

 \author{\bf V.A. Sheminova}
 \date{}

 \maketitle
 \thanks{}
\begin{center}
{Main Astronomical Observatory, National Academy of Sciences of
Ukraine
\\ Zabolotnoho 27, 03689 Kyiv, Ukraine\\ E-mail: shem@mao.kiev.ua}
\end{center}

 \begin{abstract}
Parameters are proposed for measuring the sensitivity of Fraunhofer lines to the
physical conditions in the solar atmosphere. The parameters are calculated based on
depression response functions in the LTE approximation. The sensitivity of lines to
the temperature, gas pressure, and microturbulent velocity depending on the line
and atomic parameters is investigated. The greatest relative temperature
sensitivity is shown by weak lines, while the greatest absolute sensitivity is
displayed by moderate lines of abundant heavy atoms with low ionization and
excitation potentials. The excitation potential and line strength are the crucial
factors for the temperature sensitivity. The highest pressure sensitivity is
observed for moderate lines of light atoms with very high excitation potentials
(exceeding 6 eV), and strong photospheric lines (8\,pm\,${<}W{<}$14\,pm) of heavy
atoms are the most responsive to the microturbulent velocity. The sensitivity
parameters can be also used to advantage for physical diagnostics of the
photosphere when the temperature, pressure, and microturbulent velocity
fluctuations are no more than 8\%, 50\%, and 100\%, respectively.

\end{abstract}
%-------------------------------------------------

\section{Introduction}
     \label{S-Introduction}

Local changes in the physical conditions in the solar atmosphere can cause
substantial changes in the parameters of observed absorption lines. In such a
situation, the spectral line is said to respond to the structural inhomogeneities
which have appeared in the line formation region, and the stronger is the response,
the more sensitive is the line to changes in the medium. Comprehensive information
on the sensitivity of lines to different atmospheric parameters is essential for
solving many problems in the spectrum analysis. Caccin et al. made a fundamental
contribution to the problem -- using the sensitivity function \cite{Caccin73} and the
sensitivity coefficients \cite{Caccin76}, they evaluated the changes in the central depth,
halfwidth, and equivalent width of several spectral lines caused by changes in the
temperature and microturbulent velocity. Unfortunately, the authors have not
analyzed the line sensitivity measure introduced by them, and the sensitivity
coefficients did not find wide use because of apparently cumbersome calculations.
When the sensitivity coefficients are calculated, a~variation in a model
atmosphere parameter requires that the whole model be recalculated. Besides that it
is difficult to measure the line sensitivity, since the atmospheric parameters have
a combined effect on the line profile and it is not simple to determine the
sensitivity to one of them.

Studying the contribution and response functions, we have concluded that it is
possible to determine the line sensitivity through the use of the response
functions by calculating a set of sensitivity parameters. This way is much more
handy as regards calculations though it also needs much machine time.

The prime objective of this study is to develop a technique for calculating the
sensitivity parameters based on the response functions, investigate the sensitivity
of lines to different atmospheric parameters, and find a way to separating those
lines which are highly sensitive to one of the atmospheric parameters and have a
low sensitivity to other parameters.

The first section describes the algorithm for calculating the sensitivity
parameters for absorption lines which form under the LTE conditions; the second
section presents the results for Fraunhofer lines; the third section deals with the
scope for application of the sensitivity parameters to the diagnostics of the
photosphere; and in conclusion we give principal inferences and recommendations for
choosing the necessary lines.

\section{Procedure for calculating the sensitivity \\ parameters.
}

The calculation of line sensitivity parameters is based on the idea of response
functions, which has been developed well enough by many authors
\cite{Badalyan,Beckers,Caccin77,Mein,Orrall}. The literature on the response
functions was briefly reviewed by Demidov \cite{Demidov}. Recall that the response
function describes the magnitude of response of a line to a local change in an
atmospheric parameter at each point in the atmosphere along the line of sight. It
depends on the kind of disturbance and is a function of the wavelength and optical
depth. It is called the emission or the depression response function depending on
whether it is calculated for an emission or a depression. The function maxima point
to the region in the atmosphere where the emerging emission or depression in the
line is the most sensitive to variations in the atmospheric parameters.

To estimate the sensitivity of absorption lines, we have chosen the depression
response functions proposed for the first time in \cite{Magain}. A variation in the observed
line depression, $\delta R$, which results from a local disturbance of some
atmospheric parameter is determined as \cite{Magain}

\begin{equation} \delta R ( \Delta \lambda ) = \int_{ -\infty }^{ + \infty } R F_{ R,\beta
} ( x,  \Delta \lambda )\frac{ \delta \beta (x)}  { \beta (x) }dx.\end{equation}
 Here $R= ( I_ c-I_l ) {/} I_ c $ is the line depression;
$RF_{R,\beta}$ is the response function of the depression; $\beta$ is the
atmospheric parameter which undergoes the disturbance $ \delta \beta {/} \beta$;
$x$ is the optical depth in the logarithmic scale ($ x=\log \tau_5$ where $ \tau_5$
is the optical depth at the wavelength $\lambda = 500$ nm); $ \Delta \lambda$ is
the distance to the line center. If $ \delta \beta {/} \beta$ is not zero in a thin
layer $ \delta x$ in the atmosphere close to the depth $x$, the following
approximate equality is true for the depression response function:

\begin{equation} R F_{ R, \beta }( x,\Delta \lambda ) \delta x \approx  \frac{ \delta R ( x, \Delta
\lambda )} { \delta \beta ( x ) {/} \beta ( x ) }.\end{equation}

It follows from (1) and (2) that the response function defines the rate of
variation in the line depression $ R ( x, \Delta \lambda )$  at every point $x$ in
the atmosphere relative to the rate of variation in the parameter $\beta$ which
undergoes the local disturbance $ \delta \beta {/} \beta$. If we integrate the
response function with respect to $x$ and divide it by the value of the
``emerging'' depression $R(\Delta\lambda)$, we obtain a dimensionless
quantity which is a variation in the observed depression, $\delta R{/}R$, relative
to the local variation $ \delta \beta {/} \beta$. We denote this quantity by $P_{R,
\beta}$ and call it the parameter, or the indicator, of the line depression
sensitivity. It is calculated with the expression

\begin{equation} P_ { R,\beta } ( \Delta \lambda ) =  \frac{1} {  R ( \Delta \lambda )} \int_
{ - \infty }^{+ \infty } R F_ { R, \beta }( x, \Delta \lambda ) dx .\end{equation}

As $ \delta \beta {/}\beta $ in (1) is always taken positive, the sign of the
sensitivity parameter corresponds to the sign of $\delta R (\Delta\lambda)$. It
should be remembered therefore: if the sensitivity parameter $P_{R,\beta}$ is
positive, the depression $R$ is growing when $\beta$ increases by $\delta \beta$,
and if $P_{R,\beta}$ is negative, $R$ is diminishing. By analogy with (3), we may
define the sensitivity indicator for the line equivalent width:

\begin{equation} P_ { W, \beta }  = \frac{1} {  W }  \int_ { - \infty }^ { + \infty}  R F_ { W ,\beta }( x )
dx,\end{equation}
 where $RF_{W,\beta}$ is the integral response function.
According to [13], it has the form

\begin{equation} RF_ {W ,\beta }( x )=\int_{ line } R F_ { R ,\beta }( x , \Delta
\lambda ) d  ( \Delta \lambda )    .\end{equation}

To calculate the sensitivity parameters with (3), (4), we have first to calculate
reliably the depression response functions. For the LTE photospheric lines, the
expression for the depression response function proposed in \cite{Magain} may be used:

\begin{equation} R F_{ R , \beta } ( x , \Delta \lambda ) =\beta \mu^ { -1} \ln
10\, \tau_5 \frac{\kappa_R} {\kappa_5 } \left [ {\frac{dS_R}  {d \beta}- \frac{1}
{\kappa_R} ( R-S_R ) \frac{d \kappa_R }  {d \beta} } \right ] \exp \left ( -{\frac{
\tau_ R}  {\mu} }  \right ) ,\end{equation}
 where

\[\kappa_R=\kappa_l+\kappa_c \frac{B} {I_c} ; ~~~S_R=\left (
{1- \frac{B} {I_c} }  \right ) { / } \left({1+ \frac{\kappa_c} {\kappa_l} \frac{B}
{I_ c} }  \right ); ~~R=1-\frac{I_l }{I_c };\]

\[ I_{c,l} ( \tau_ {c,l} )=\exp \left ( \frac{ \tau_{c,l}} {\mu } \right )
\int_{\tau_{c ,l}}^{\infty} \frac{B} {\mu} \exp \left ( -\frac{t} {\mu } \right
)dt;~~ \tau_R=\tau_l+ \int_0^{\tau_c} \frac{B} {I_c} dt .\]

 Here $\mu = \cos \theta$; $k_l$, $ k_c$ and $\tau_l$, $\tau_c$ are the absorption
coefficients and optical depths in the line ($l$) and in the continuum ($c$); $B$
is the Planck function.

Since the response function depends on the kind of disturbance, i.e., on $\beta$,
it is necessary to calculate the response function for each atmospheric parameter
whose effect on the line profile we want to estimate. Let the temperature variation
$\delta T/T$ occur in the solar atmosphere, then the depression function of the
response to this temperature variation, according to (6), is

\begin{equation} R F_{ R ,T } ( x , \Delta \lambda ) =T \mu^{-1} \ln10\, \tau_5 \frac{\kappa_R} {\kappa_5 }
\left [ {\frac{dS_R} {d T}-\frac{1}  {\kappa_R} ( R-S_R ) \frac{d \kappa_ R } {d T}
}  \right ] \exp \left ( { -\frac{\tau_ R} {\mu} }  \right ) .\end{equation}

Similarly we get the expressions for the depression functions of the response to
the variations in the gas pressure, $\delta P/P$,

\begin{equation} R F_ { R ,P } ( x , \Delta \lambda ) =P \mu^{-1} \ln 10\, \tau_5
\frac{\kappa_ R}{\kappa_5 } \left [ {\frac{dS_R} {d P}-\frac{1} {\kappa_ R} ( R-S_R
) \frac{d \kappa_R } {d P} }  \right ] \exp \left ( { -\frac{\tau_R} {\mu} } \right
),\end{equation}
 and in the microturbulent velocity, $\delta V/V$,

\[ R F_{ R ,V } ( x , \Delta \lambda ) =- V^2  \mu^{-1} \ln
10\, \tau_5 \frac{\kappa_l} {\kappa_5 } \left ( {\frac{2 RT} {m} +V^2 } \right )^{-1}
\left ( {\frac{\Delta \lambda} {\Delta \lambda_D H ( a,v )}\frac{d [ H ( a,v ) ] }
{dv}+1} \right)\times \]

\begin{equation}\times \left [ {\frac{\kappa_c}  {\kappa_ R}  \left ( {1-\frac{B}  {I_c}
} \right ) \frac{B}  {I_ c} -  ( R-S_R ) }  \right ]  \exp \left ( -\frac{\tau_R}
 {\mu}   \right ).\end{equation}

The algorithm for calculating the sensitivity parameters of the Fraunhofer lines is
based on formulae (7)--(9). The calculation program is based on the SPANSAT program
for the Fraunhofer line calculations \cite{Gadun88}.

\section{Calculation and analysis of sensitivity parameters
}

It is well known that the temperature sensitivity of a line depends on the line
excitation potential, the sensitivity decreasing with increasing potential. Just
this rule is commonly used by observers when they choose lines. However no
investigations have thus far been carried out concerning the line sensitivity to
other atmospheric parameters and the dependence of the sensitivity on the line
atomic parameters such as the wavelength $\lambda$, ionization potential, atomic
weight, as well as on the abundance and on the line central depth, equivalent
width, and half-width. The employment of the above technique for calculating the
line sensitivity parameters allowed us to elucidate the problem.

\subsection{Starting data for calculations }

%___________________________________ Table 1
%
{\footnotesize
 \begin{table}[t] \centering
 \parbox[b]{14cm}{
\caption{ Response of Line Central Depths to Gas Pressure Disturbance. }
 \label{T:1}
\vspace{0.3cm}}
 \footnotesize
\begin{tabular}{clrrrrrrrrr}
 \hline

no.& $\lambda$,\,nm& $AM$& $A$& $IP$,\,eV&$EP$,\,eV&$R$& $\log\tau_{5,R}$&
$P_{R,T}$& $P_{R,V}$&$P_{R,P}$ \\
 \hline

1&477.588~ C I& 12.01& 8.65& 11.26~& 7.49& 0.138& -0.274& ~~2.24& -0.09& -1.08    \\
2&658.762~ C I& 12.01& 8.65& 11.26~& 8.53& 0.083& -0.329& ~~3.44& -0.09& -1.35      \\
3&777.539~ O I& 16.00& 8.90& 13.62~& 9.14& 0.249& -0.336& ~~4.06& -0.11& -1.36      \\
4&514.884~ Na I& 22.99& 6.32& 5.14& 2.10& 0.128& -0.884& -6.19& -0.19& -0.13        \\
5&669.603~ Al I& 26.98& 6.49& 5.99& 3.14& 0.263& -1.128& -5.09& -0.14& -0.29        \\
6&462.736~ Si I& 28.09& 7.64& 8.15& 5.08& 0.173& -0.845& -2.94& -0.20& -0.31        \\
7&613.185~ Si I& 28.09& 7.64& 8.15& 5.61& 0.183& -0.928& -2.57& -0.18& -0.49        \\
8&674.163~ Si I& 28.09& 7.64& 8.15& 5.98& 0.101& -0.891& -2.28& -0.22& -0.60        \\
9&567.181~ Sc I& 44.96& 3.06& 6.56& 1.45& 0.110& -1.254& -10.18~& -0.28& ~0.04       \\
10&623.936~ Sc I& 44.96& 3.06& 6.56& 0.00& 0.063& -1.424& -14.70~& -0.31& ~0.15     \\
11&477.826~ Ti I& 47.90& 5.06& 6.82& 2.24& 0.183& -1.139& -8.35& -0.28& ~0.06       \\
12&546.049~ Ti I& 47.90& 5.06& 6.82& 0.05& 0.107& -1.420& -14.72~& -0.32& ~0.22     \\
13&577.403~ Ti I& 47.90& 5.06& 6.82& 3.30& 0.101& -1.021& -7.62& -0.31& -0.10       \\
14&592.212~ Ti I& 47.90& 5.06& 6.82& 1.05& 0.200& -1.404& -10.55~& -0.26& ~0.03     \\
15&445.777~ V I& 50.94& 4.00& 6.74& 1.87& 0.119& -1.143& -9.10& -0.31& ~0.15        \\
16&482.745~ V I& 50.94& 4.00& 6.74& 0.04& 0.127& -1.420& -13.44~& -0.31& ~0.25      \\
17&521.412~ Cr I& 52.00& 5.64& 6.77& 3.37& 0.204& -1.052& -6.66& -0.29& -0.09       \\
18&523.896~ Cr I& 52.00& 5.64& 6.77& 2.71& 0.189& -1.122& -7.87& -0.29& -0.02       \\
19&666.108~ Cr I& 52.00& 5.64& 6.77& 4.19& 0.110& -1.010& -6.22& -0.32& -0.27       \\
20&557.702~ Fe I& 55.85& 7.64& 7.87& 5.03& 0.118& -0.944& -5.35& -0.33& -0.26       \\
21&561.135~ Fe I& 55.85& 7.64& 7.87& 3.63& 0.101& -1.077& -7.93& -0.35& -0.07       \\
21&568.024~ Fe I& 55.85& 7.64& 7.87& 4.19& 0.112&  -1.025& -6.81& -0.33& -0.14      \\
22&662.502~ Fe I& 55.85& 7.64& 7.87& 1.01& 0.149& -1.521& -12.78~& -0.31& ~0.03     \\
23&673.952~ Fe I& 55.85& 7.64& 7.87& 1.56& 0.108& -1.427& -12.08~& -0.33& ~0.01     \\
24&697.194~ Fe I& 55.85& 7.64& 7.87& 3.02& 0.113& -1.243& -9.25& -0.33& -0.11       \\
25&401.109~ Co I& 58.93& 4.92& 7.86& 0.10& 0.080& -1.454& -14.33~& -0.37& ~0.34     \\
26&459.464~ Co I& 58.93& 4.92& 7.86& 3.62& 0.112& -1.015& -6.76& -0.34& -0.01       \\
27&514.979~ Co I& 58.93& 4.92& 7.86& 1.73& 0.102& -1.313& -10.92~& -0.35& ~0.09     \\
28&611.977~ Ni I& 58.70& 6.22& 7.63& 4.26& 0.107& -1.010& -5.51& -0.34& -0.27       \\
29&522.007~ Cu I& 63.55& 4.10& 7.73& 3.82& 0.158& -1.006& -5.73& -0.35& -0.16       \\
30&636.235~ Zn I& 65.38& 4.60& 9.39& 5.79& 0.192& -0.927& -0.39& -0.25& -0.63       \\
31&412.830~ Y I& 88.91& 2.24& 6.22& 0.07& 0.120& -1.272& -12.52~& -0.45& ~0.34      \\
32&468.780~ Zr I& 91.22& 2.56& 6.84& 0.73& 0.136& -1.351& -12.13~& -0.43& ~0.23     \\
    \hline
 \end{tabular}
 \end{table}}
 \noindent
 %%%%%%%%%%%%%%%%%%%%%%%%%%%%%%%%%%%%%%%%%%%%%%%%%%%%% Tab1 cont
{\footnotesize
 \begin{table}[t] \centering
 \parbox[b]{14cm}{
%\caption[]
{\hspace{10.cm} Table 1. (continued) }
 \label{T:1}
\vspace{0.3cm}}
 \footnotesize
\begin{tabular}{ccrrrrrrrrr}
 \hline

no.&$\Delta\lambda_{R/2}$,\,pm & $\log \tau_{5,R/2}$ & $P_{R/2,}T$ & $P_{R/2,V}$
&$P_{R/2,P}$& $W,\,pm$ & $\log\tau_{5,W}$ & $P_{W,T}$&  $P_{W,V}$& $P_{W,P}$\\
 \hline

 1&5.1& -0.147& ~2.91& 0.05& -1.13& 1.90& -0.196& ~2.65& 0.01& -1.09        \\
 2&7.2& -0.197& ~3.65& 0.03& -1.38& 1.70& -0.248& ~3.60& -0.01~& -1.35        \\
 3&7.2& -0.207& ~4.50& 0.09& -1.45& 2.10& -0.260& ~4.35& 0.01& -1.40          \\
 4&3.7& -0.749& -5.94& 0.05& -0.10& 1.20& -0.793& -6.00& -0.03~& -0.10        \\
 5&5.0& -0.868& -5.58& 0.14& -0.23& 3.57& -0.966& -5.35& 0.05& -0.23          \\
 6&3.4& -0.645& -2.92& 0.06& -0.24& 1.95& -0.696& -2.90& -0.01~& -0.24        \\
 7&4.6& -0.706& -2.71& 0.08& -0.43& 2.57& -0.771& -2.70& 0.02& -0.41          \\
 8&5.3& -0.684& -2.35& 0.01& -0.49& 1.65& -0.735& -2.30& -0.04~& -0.50        \\
 9&3.0& -1.102& -10.73~& 0.17& ~0.08& 1.37& -1.159& -10.45~& 0.03& ~0.07       \\
10&3.1& -1.304& -14.89~& 0.14& ~0.16& 0.71& -1.352& -14.80~& -0.03~& ~0.16    \\
11&2.6& -0.973& -8.85& 0.19& ~0.11& 1.57& -1.037& -8.60& 0.03& ~0.10          \\
12&2.7& -1.295& -14.94~& 0.15& ~0.23& 0.78& -1.341& -14.80~& 0.00& ~0.22      \\
13&3.1& -0.888& -7.66& 0.14& -0.07& 1.02& -0.934& -7.60& 0.00& -0.07          \\
14&3.1& -1.214& -11.72~& 0.20& ~0.08& 1.06& -1.291& -11.15~& 0.05& ~0.06      \\
15&2.3& -0.995& -9.45& 0.18& ~0.19& 1.19& -1.052& -9.25& 0.01& ~0.17          \\
16&2.4& -1.270& -14.07~& 0.17& ~0.28& 1.17& -1.330& -13.75~& 0.00& ~0.26      \\
17&2.9& -0.869& -7.01& 0.18& -0.04& 1.65& -0.934& -6.85& 0.03& -0.04          \\
18&2.8& -0.956& -8.34& 0.19& ~0.02& 1.52& -1.016& -8.10& 0.04& ~0.02          \\
19&3.7& -0.840& -6.28& 0.12& -0.20& 1.23& -0.893& -6.25& -0.01~& -0.20        \\
20&3.1& -0.768& -5.39& 0.13& -0.19& 1.21& -0.819& -5.35& -0.01~& -0.19        \\
21&2.8& -0.948& -8.02& 0.17& -0.05& 0.90& -0.995& -7.95& -0.01~& -0.05        \\
21&2.9& -0.881& -6.94& 0.17& -0.11& 1.11& -0.930& -6.85& 0.01& -0.11          \\
22&3.2& -1.349& -13.64~& 0.19& ~0.06& 1.39& -1.417& -13.25~& 0.02& ~0.05      \\
23&3.2& -1.276& -12.60~& 0.18& ~0.03& 1.11& -1.334& -12.35~& 0.01& ~0.02      \\
24&3.4& -1.096& -9.61& 0.18& -0.09& 1.00& -1.154& -9.45& -0.00~& -0.09        \\
25&1.8& -1.319& -14.42~& 0.17& ~0.35& 0.52& -1.370& -14.35~& -0.01~& ~0.34    \\
26&2.3& -0.872& -6.91& 0.23& ~0.02& 1.13& -0.926& -6.85& 0.02& ~0.01          \\
27&2.4& -1.171& -11.26~& 0.20& ~0.10& 0.94& -1.227& -11.10~& 0.00& ~0.10      \\
28&3.1& -0.860& -5.66& 0.22& -0.24& 1.15& -0.913& -5.60& 0.02& -0.24          \\
29&2.6& -0.852& -5.93& 0.27& -0.13& 1.31& -0.908& -5.85& 0.04& -0.14          \\
30&3.6& -0.683& -0.45& 0.38& -0.63& 2.42& -0.781& -0.50& 0.12& -0.60          \\
31&1.7& -1.131& -12.69~& 0.24& ~0.35& 0.83& -1.184& -12.60~& 0.01& ~0.35      \\
32&2.0& -1.180& -12.76~& 0.27& ~0.26& 0.97& -1.245& -12.45~& 0.05& ~0.25      \\
    \hline
 \end{tabular}
 \end{table}}
 \noindent
 %%%%%%%%%%%%%%%%%%%%%%%%%%%%%%%%%%%%%%%%%%%%%%%%%%%%%

We have chosen ten groups of lines from the line list given in \cite{Gurtovenko89}, so that we
might deduce the dependences we are interested in. The first group contains lines
of different elements with different excitation potentials from 0 to 9 eV and with
central depths close to each other in the range from 0.1 to 0.2. Table 1 gives the
starting data for the first-group lines; $AM$ is the atom mass, $A$ is the element
abundance, $IP$ is the ionization potential, $EP$ is the excitation potential; $R$,
$\Delta\lambda_{R/2}$, $W$ are the central depth, half-width, and equivalent width
of the line; $\log \tau_{5,R}$, $\log\tau_{5,R/2}$, $\log\tau_{5,W}$ are the
effective optical depths of formation of the line center, of the part of the line
profile which corresponds to the half-width, and of the entire line on the average.
The results of calculations for the lines in the first group permitted us to find
how the line sensitivity parameters depend on the excitation potential, ionization
potential, abundance, and atomic weight. The second group comprises the iron lines
for finding the dependence of sensitivity parameters on the line wavelength. We
have selected five lines with $EP=4$ eV, $R=0.65$. The remaining eight groups of
lines, which include the Fe I lines only, served for determining the dependence of
line sensitivity on the line central depth, half-width, and equivalent width. In
each of these eight groups we selected lines with close excitation potentials,
namely: 0, 1, 1.5, 2.5, 3, 3.5, 4, 5 eV, with different central depths from 0.05 to
0.8 and equivalent widths from 0.4 to 13 pm. Such a selection of lines permitted a
simultaneous investigation of the dependence of line sensitivity on both the line
strength and the excitation potential, excluding the effect of the ionization
potential, abundance, and atomic weight; we could also obtain the sensitivity
indicators for actual lines in the solar spectrum.

We used the model atmosphere HOLMU \cite{Holweger} in the calculations of sensitivity
parameters. The data on the oscillator strengths and microturbulent velocity, which
varies with height in the photosphere, were taken from \cite{Gurtovenko89}. The van der Waals
damping constant was taken with a correction factor of 1.5. The macroturbulent
velocity was ignored, since it does not enter the response function calculations.
The results of calculations are shown in the figures, and the results for the lines
of the first group are given also in Table 1. The sensitivity parameters for the
complete list of unblended lines \cite{Rutten} are likely to be published separately in
tabulated form.

\subsection{Temperature sensitivity of lines
 }

We consider changes in the temperature sensitivity of lines as functions of
principal atomic parameters. The calculations for lines in the first group confirm
that the temperature sensitivity depends on the excitation potential. This is
clearly seen in Fig.~1a, where the relation $P_{R,T}(EP)$ is shown. The dependence
of the temperature sensitivity on the sum of excitation and ionization potentials
$EP +IP$ is similar in general terms to the dependence of $P_{R,T}$ on $EP$.
Analyzing the data in Fig.~1a and in Table~1, we can draw the following
conclusions. Lines of atoms with high abundance and large atomic weight, with low
excitation and ionization potentials are the most responsive to temperature. As
seen in the figure, the excitation potential of the lower level has a dominant role
in the temperature sensitivity of the line. Lines with excitation potentials to 2
eV are highly responsive, while those with $EP$ from $\approx 5.5$ to $\approx7.5$
eV are the least responsive. The sensitivity increases again with a further
increase of $EP$. It is intriguing that the sign of the temperature sensitivity
parameter changes. Absorption lines of light atoms with $EP\approx$ 6--7 eV become
stronger with growing temperature, i.e., their central depths and equivalent widths
increase. It should be remarked that in this case their temperature sensitivity,
while increasing, still remains low. For instance, for one and the same change in
the temperature, lines with $EP \approx 10$ eV change similarly to lines with $EP
\approx5.5$~eV, though with the opposite sign.

%------------------------------------------------------------- Fig1
\begin{figure}
   \centering
   \includegraphics[width=13 cm]{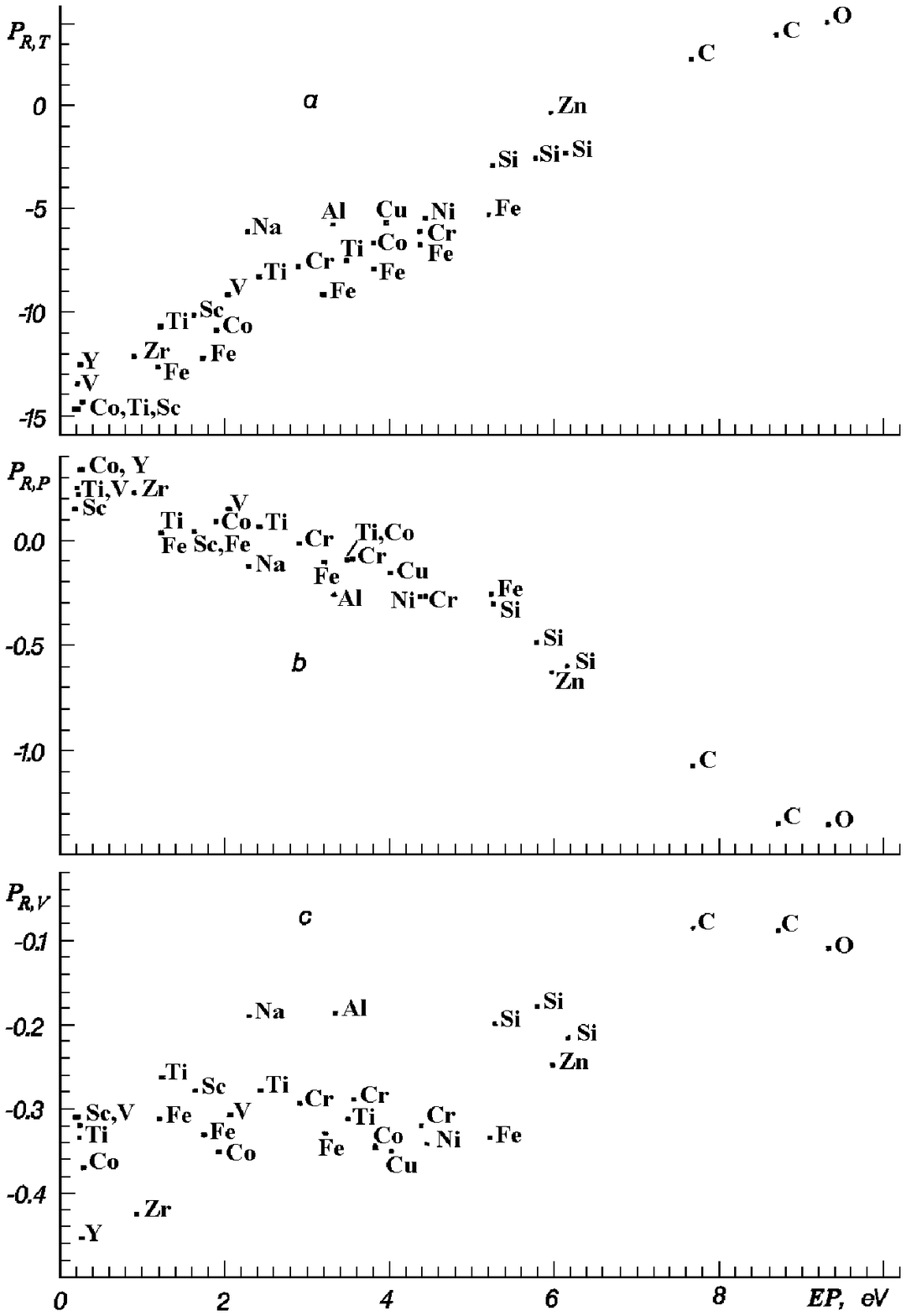}
%   \hfill
%\parbox[b]{10.5cm}{ \vspace{0.0cm}
   \caption[]{Parameters of sensitivity of the line central depth to temperature
(a), gas pressure (b), and microturbulent velocity (c) as functions of the
excitation potential.
 } \label{Fig1}
 %}

\end{figure}
%____________________________________________________________

%------------------------------------------------------------- Fig2
\begin{figure}[t]
   \centering
   \includegraphics[width=7  cm]{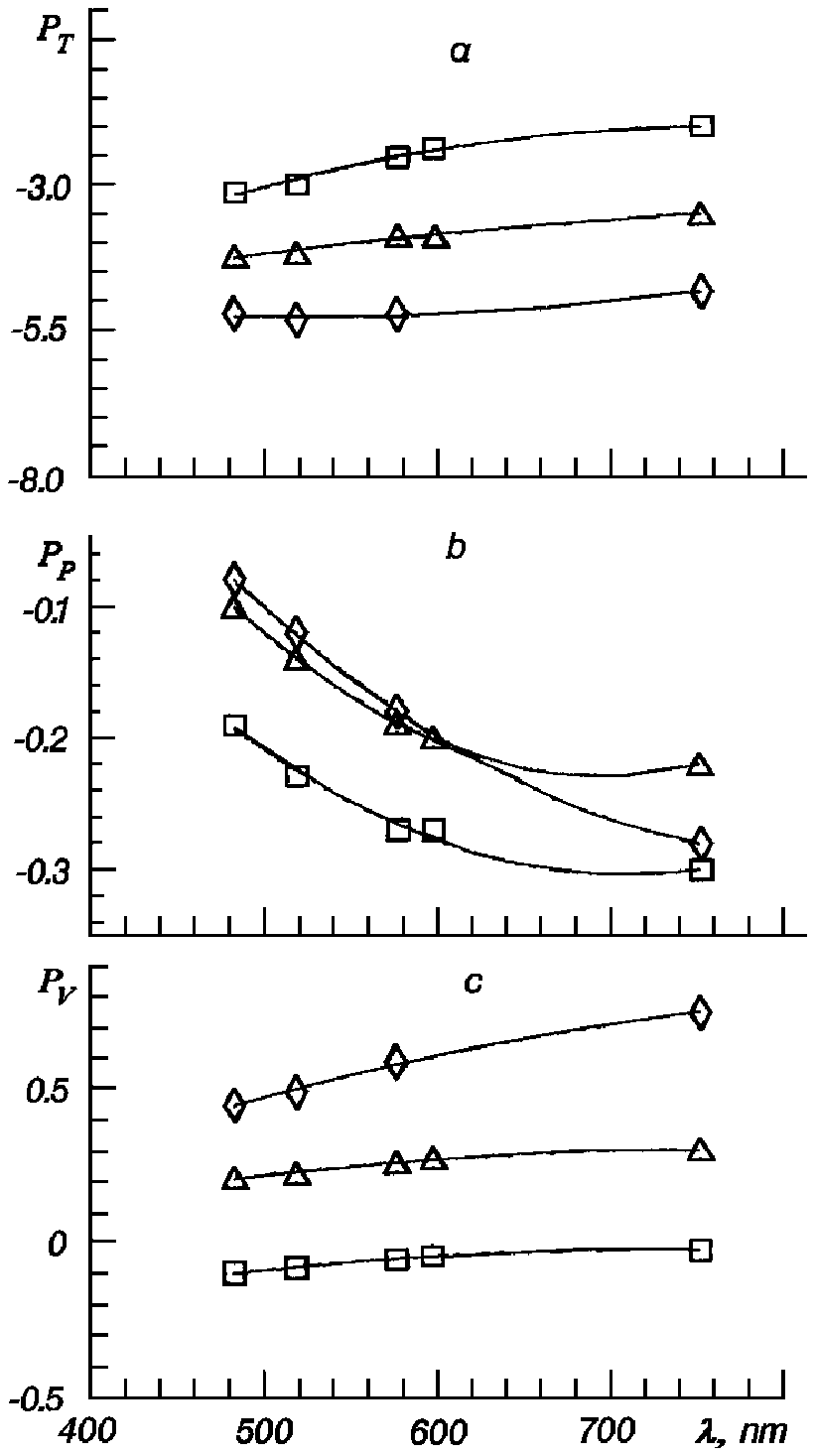}
   \hfill
\parbox[b]{6.5cm}{ \vspace{0.0cm}
   \caption[]{ Parameters of sensitivity of the line central
depth (squares), half-width depth (diamonds), and equivalent width (triangles) to
temperature (a), gas pressure (b), and microturbulent velocity (c ) as functions of
wavelength.

 } \label{Fig2}
 }

\end{figure}
%____________________________________________________________

The wavelength dependence of the temperature sensitivity is shown in Fig.~2a. The
sensitivity grows with decreasing wavelength, i.e., there is a tendency for the
line temperature sensitivity to grow in the shortwave region and to diminish for
longer wavelengths.

The sensitivity parameters for three line characteristics -- the central depth,
half-width depth, and equivalent width -- differ slightly in the case of weak
lines. This can be seen from Table 1. The most temperature-sensitive is the
half-width depth, less sensitive is the equivalent width, and, finally, the central
depth. Now we analyze the dependence of the line sensitivity on the central depth
$R$, equivalent width $W$, and half-width $ \Delta\lambda_ {R/2}$. Towards this
end, we have plotted the relationships $P_{R,T}$---$R$ (Fig.~3a), $P_{W,T}$---$W$
(Fig.~3b), and $P_{R/2,T}$---$\Delta\lambda_{R/2}$ (Fig.~3c) based on the
calculations of sensitivity parameters for lines in eight groups. Since the
sensitivity depends strongly on $EP$, we obtained a family of curves, each curve
representing the dependence of the sensitivity parameter on $R$, or $W$, or
$\Delta\lambda_{R/2}$ for a certain value of $EP$ in accordance with the
calculation results. The figures show the curves for $EP$ = 0, 1, 3, 4, 5 eV only.
As it is impossible to choose actual lines in the solar spectrum for each group
with the same wavelengths, a small scatter of points can be seen in all figures,
which is due to the wavelength dependence of the sensitivity parameters. It is
evident from the figures that easily excited weak lines are the most
temperature-sensitive. Their sensitivity decreases with increasing $R$ and $W$.
Strong lines with high excitation potentials have low sensitivity. The dependence
of the sensitivity on the half-width indicates that lines become less sensitive
with growing half-width, i.e., narrow lines are more sensitive to the temperature.

Thus, the temperature sensitivity of lines depends the most strongly on the
excitation potential above all and to a lesser degree on the line strength. Using
the same analysis procedure as for the temperature sensitivity, we studied the
sensitivity of lines to the gas pressure and microturbulent velocity.

\subsection{Sensitivity of lines to the gas pressure  }

A decisive role in the pressure sensitivity of lines belongs to the low excitation
potential. The run of the dependence of line sensitivity to gas pressure on the
excitation potential (Fig.~1b) is opposite to that of the sensitivity to
temperature. Lines with $EP$ from 0 to 5 eV have a very low sensitivity to
pressure. From 0 to 2 eV, the sensitivity decreases, the central depths and
equivalent widths of lines of heavy atoms slightly increasing with pressure.
Beginning with 2 eV, lines with higher excitation and ionization potentials and
lines belonging to less abundant and lighter atoms become more sensitive to
pressure. The most pressure-sensitive lines are those with $EP> 6$ eV. These may be
the lines of carbon, oxygen, nitrogen, as well as lines of ions. Lines of atoms in
the iron group may be considered as having low sensitivity to pressure (see Fig.~1b
and Table 1). The wavelength also affects the sensitivity of lines to pressure
fluctuations. As evident from Fig.~2b, the sensitivity increases with the
wavelength, i.e., the effect is opposite to the temperature sensitivity.

The pressure sensitivity depends on the line parameters in a more complicated way
than the temperature sensitivity (Fig.~4). A stronger dependence of the pressure
sensitivity on $\lambda$ increases the spread of points on each plot for a certain
$EP$. The sensitivity grows at first with $R$ and $W$ and then diminishes. For
instance, the greatest sensitivity of the central part of Fe I lines (Fig.~4a) is
observed in those lines for which $R = 0.4$--0.6 and $EP = 5$~eV. When $EP$ becomes
smaller, the maximum sensitivity shifts slightly towards stronger lines. As regards
the dependence of the sensitivity on $W$ (Fig.~4b), the pattern is the same in
general. For example, the greatest sensitivity for lines with $EP = 4$~eV falls
within the range of $W$ from 5 to 7 pm. We can say also that the sensitivity grows
at first in the main with the line half-width and then diminishes (Fig.~4c). In
contrast to the temperature sensitivity, narrow lines are less sensitive to
pressure.

Thus, the greatest response to the gas pressure fluctuations is observed in
moderate lines with very high excitation potentials.

%------------------------------------------------------------- Fig3
\begin{figure}
   \centering
   \includegraphics[width=12. cm]{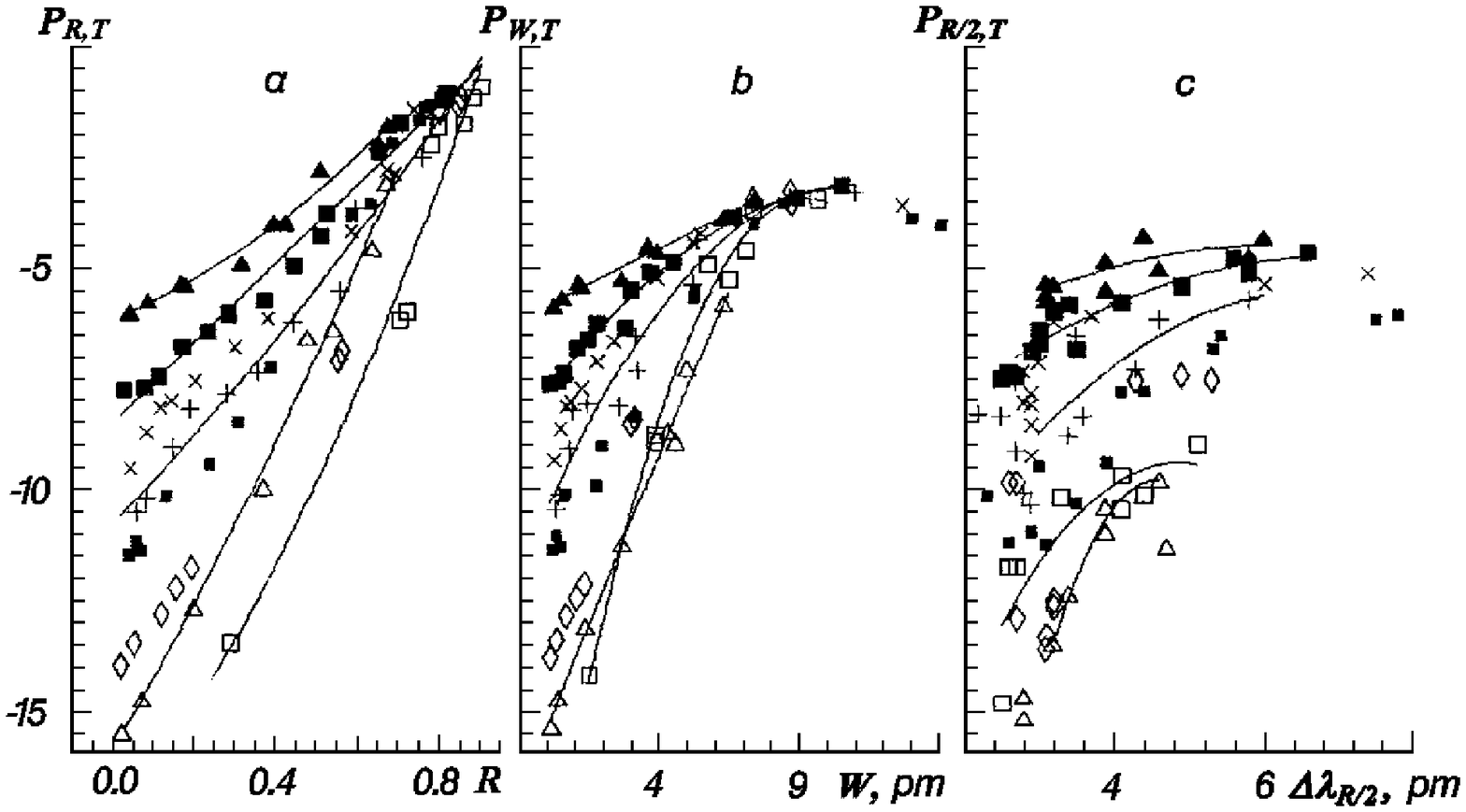}
%   \hfill
%\parbox[b]{10.5cm}{ \vspace{0.0cm}
   \caption[]{ Parameters of sensitivity  to  temperature for
central depth (a), equivalent width (b), and half-width depth (c) as functions of
the parameters $R$, $W$, and $\Delta\lambda_{R/2}$. $EP=0$ (light squares), 1
(light triangles), 1.5 (diamonds), 2.5 (dark squares), 3 (plusses), 3.5 (crosses),
4 (big dark squares), 5 (dark triangles). The curves  show the dependencies for the
lines with $EP$ = 0, 1, 3, 4, 5 eV. } \vspace{0.3cm}
% } \label{Fig3}
% %}

%\end{figure}
%____________________________________________________________

%------------------------------------------------------------- Fig4
%\begin{figure}
%   \centering
   \includegraphics[width=12. cm]{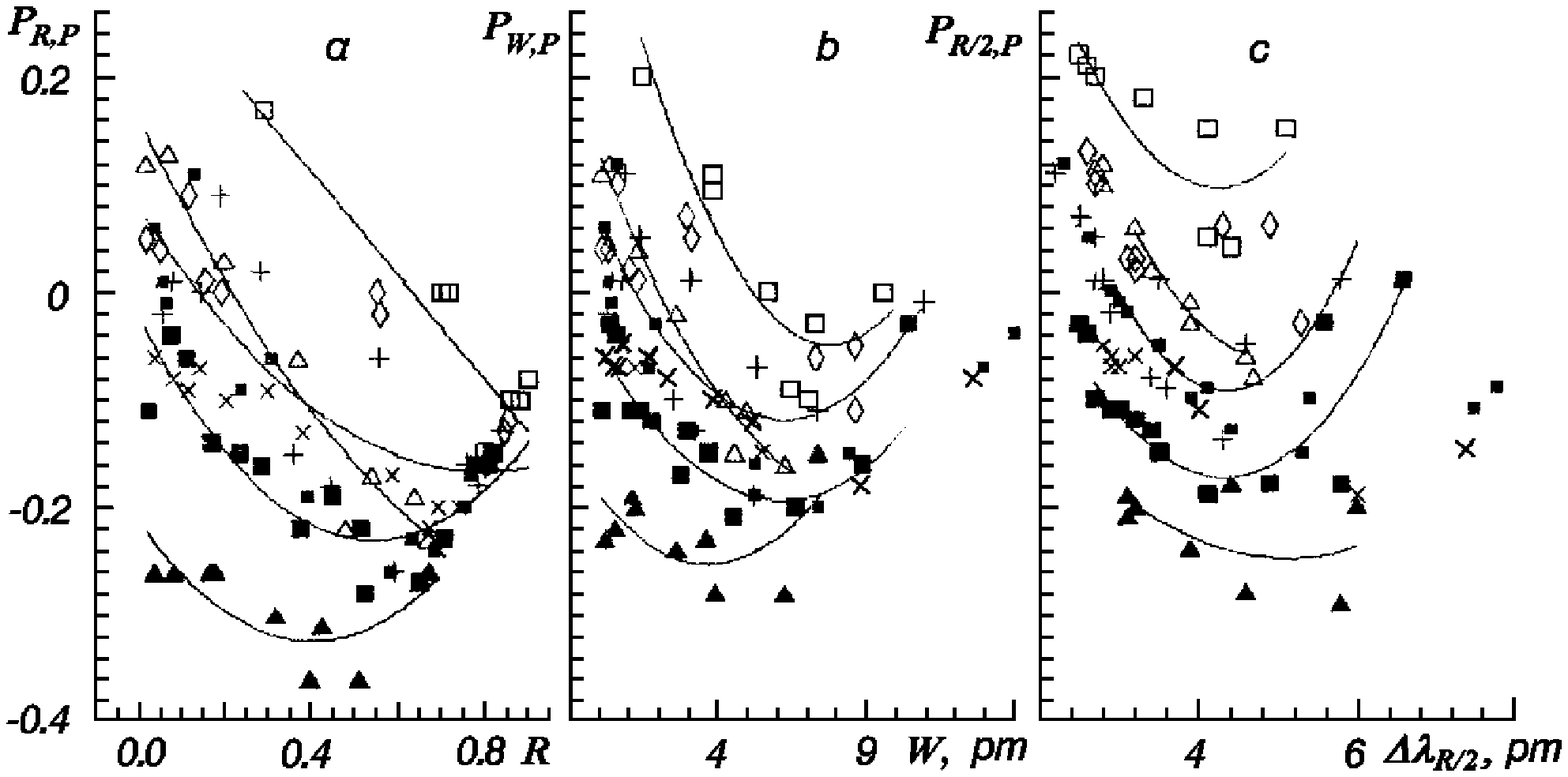}
%    \includegraphics[width=11. cm]{sh5.eps}

%   \hfill
%\parbox[b]{10.5cm}{ \vspace{0.0cm}
   \caption[]{ Parameters of
sensitivity to the gas pressure. Format is the same as in Fig. 3.} \vspace{0.3cm}
% } \label{Fig4}
% %}
%\end{figure}
%%____________________________________________________________
%\noindent
%%------------------------------------------------------------- Fig5
%\begin{figure}
%   \centering
  \includegraphics[width=12. cm]{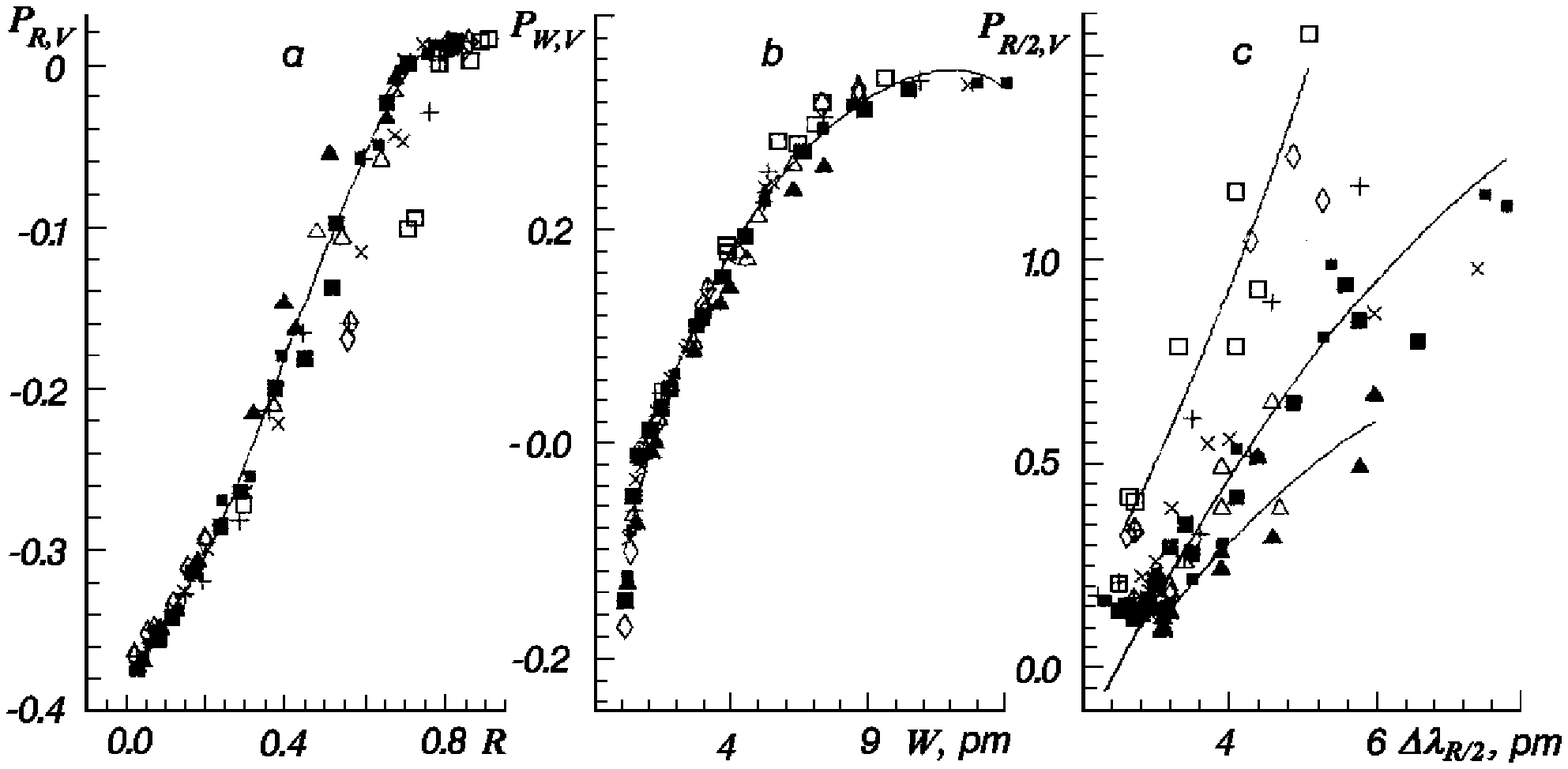}
 %  \hfill
%\parbox[b]{10.5cm}{ \vspace{0.0cm}
   \caption[]{ Parameters of sensitivity to the microturbulent velocity.
   Format is the same as in Fig. 3. The curves  show the dependencies for the
lines with  $EP= 2.5$ eV (a,b) and $EP= 0$, 2.5, 5 eV (c).
 } \label{Fig5}
% }

\end{figure}
%____________________________________________________________

\subsection{Sensitivity of lines to the microturbulent velocity }

The analysis of parameters of sensitivity to the microturbulent velocity (Table 1,
Figs 1c and 5) reveals that the sensitivity of lines to the microturbulent velocity
depends on $EP$ much more weakly than the temperature and pressure sensitivities. A
line is most responsive in the middle part of its wing and least responsive in its
central part. For the line central depth, the sensitivity to the microturbulent
velocity slightly decreases with increasing $EP$, this effect being stronger for
stronger lines. The dependence of $P_{R,V}$ on $EP$ disappears at all for very
strong lines. The the half-width depth, the sensitivity of weak lines remains
almost the same with growing $EP$, but it increases for strong lines, while for the
equivalent width the dependence of $P_{W,V}$ on $EP$ is virtually absent. Figure
1$c$ shows how the sensitivity depends on $EP$ for the central depth. The atomic
weight exerts strong control over the sensitivity parameter $P_{R,V}$, and we may
state that this effect is predominant as compared to other atomic line parameters.
It is clearly seen in Fig.~1c how four line groups separate -- lines of very heavy
atoms (Y, Zr), heavy (Sc, Ti, V, Cr, Mn, Fe, Co, Cu, Zn, Ni), medium-heavy (Na, Mg,
Al, Si), and light atoms (C, O). The heavier the atom, the higher the sensitivity
of line central part to the microturbulent velocity. The atomic weight affects the
sensitivity of the equivalent width to a lesser degree than the sensitivity of the
line central part (Table 1). The sensitivity of a line to the microturbulent
velocity grows with the line wavelength (Fig.~2c).

Figure 5 depicts the line sensitivity as a function of line parameters. The
sensitivity of the central depth to the microturbulent velocity decreases quite
strongly with increasing $R$ (Fig.~5a). Lines with $R \approx 0.7$ are practically
insensitive in their central parts. The sensitivity of the equivalent width grows
with $W$ (Fig.~5b), attaining its maximum at $W\approx 7$ pm, and remains invariant
up to $W \approx 14$ pm, and after that a fall in sensitivity is evident. Figure~5c
indicates that wider lines are more sensitive to the microturbulent velocity.

Thus, the sensitivity of a line to the microturbulent velocity is practically
independent of atomic parameters. It is governed mainly by the line equivalent
width and half-width.

\subsection{The absolute line sensitivity }

We have analyzed the sensitivity of lines, based on the sensitivity indicators
calculated with expressions (3)--(4), which characterize a relative variation in
line parameters. We call them the relative sensitivity indicators. If we multiply
expressions (3), (4) by $R$ and $W$, we can obtain the absolute sensitivity
parameters, which characterize absolute depression variations.

We analyzed the absolute sensitivity parameters together with the relative ones.
Clearly the dependence of the absolute sensitivity parameters on $R$ and $W$ is not
the same as shown in Figs~3--5. We shall comment it briefly. If we examine how the
absolute sensitivity of central intensities to variations in the atmospheric
parameters $T, P,$ and $V$ depends on $R$, we find that moderate lines are the most
sensitive, and not weak lines as it was for the relative sensitivity. The maximum
absolute temperature sensitivity does not shift for different $EP$, it only grows
with decreasing $EP$. The maximum pressure sensitivity shifts towards smaller $R$
with increasing $EP$, and what is more, it grows. The maximum sensitivity to the
microturbulent velocity shifts towards smaller $R$ and diminishes with increasing
$EP$. For the absolute sensitivity of equivalent widths, the situation is as
follows. The absolute temperature sensitivity increases with $W$, this relationship
bearing a resemblance to the curve of growth -- steeper parts of the curve refer to
weak and strong lines, while the flat part refers to moderately strong lines.
Depending on $EP$, the curve splits into several curves in the region of weak and
moderately strong lines. The curves for lines with low $EP$ lie higher. The
dependence of the absolute pressure sensitivity on $W$ has a maximum for moderate
lines which is more clear-cut than in Fig.~4b. The absolute sensitivity to the
microturbulent velocity grows linearly with $W$, beginning with $W$~=~1.5~pm. The
sensitivity does not stop growing for large values of $W$.

What sensitivity indicators are better to use for determining line sensitivity --
absolute or relative ones? This is likely to depend on the user and the specific
problem. It is not difficult to pass from one indicators to other ones. One should
know, however, that different lines will have the highest sensitivity (this is
especially true for the temperature sensitivity) depending on what sensitivity we
bear in mind -- the absolute or relative one.

\subsection{On the scope for application of the sensitivity parameters \\to
the diagnostics of the photosphere }

The sensitivity indicators may be also used to determine variations of atmospheric
parameters with height. It has been demonstrated with model response functions
\cite{Karpinskii}
that the response functions may be used in the spectrum analysis with the aim to
determine separately the parameters of the measured solar radiation intensity with
a better resolution, even though the response to a disturbance is not always linear
and adequate.

Let us consider how appropriate the sensitivity indicators proposed here are for
this purpose. First of all, it is necessary to choose correctly the lines, so that
deviations of physical parameters from their initial (model) values could be
obtained as functions of height with the help of sensitivity parameters from the
measurements of observed variations in the line depression. To this end, one has to
find lines highly sensitive to a particular atmospheric parameter and at the same
time only slightly sensitive to other parameters. Moreover, these lines should be
formed at different heights in the photosphere. It is quite easy to find deviations
(disturbances in our case) of atmospheric parameters for such a set of lines from
the calculated sensitivity parameters. The accuracy of the results lies within the
limits defined by the assumptions under which the response functions were derived
\cite{Caccin77}. Recall that the disturbance should be small and the LTE conditions should be
met in the line formation region.

We assume also that the disturbance of an atmospheric parameter occurs throughout
the region of the effective line formation. The region of formation of the
effective depression extends for photospheric lines over 260--200~km on the average
for the line center and 225--150~km for the line wing with  $R$ = 0.01 (from the data of
\cite{Sheminova92}). Let the magnitude of disturbance be a certain fraction of the
parameter. For example, if this is 5\%, then $\delta\beta = 0.05 \beta$    and $
\delta\beta /\beta = 0.05$, i.e., $\delta\beta/\beta$ = const. Then it follows from
(1):

\begin{equation}\delta R (  \Delta \lambda ) = \frac{\delta \beta } \beta  \int_ { - \infty }^
{+ \infty}  RF_ { R, \beta} (x,\Delta \lambda ) dx .\end{equation}

Under the adopted assumptions, an relative deviation of the atmospheric parameter
from its initial value can be calculated as follows:

\begin{equation}\frac{ \delta \beta } {\beta} = \frac{ \delta R ( \Delta \lambda )} { R ( \Delta
\lambda ) }\cdot \frac{1} { P_{ R, \beta } ( \Delta \lambda ) } .\end{equation}

In this case the line response $\delta\beta$ always depends linearly on the
disturbance magnitude, and there are no difficulties with the nonlinearity and
inadequacy pointed to in \cite{Karpinskii}. Now we determine the limiting values of $\delta\beta
/\beta$ within which expressions (10)--(11) remain true. We calculated directly the
central depths of lines using the initial HOLMU model and a set of ``disturbed''
HOLMU models. The following parameters were taken as independent ones for initial
models (all other model parameters were calculated later with them): the geometric
height $H$ of atmospheric layers; temperature $T(H)$; gas pressure $P(H)$;
microturbulent velocity $V(H)$ which was approximated according to \cite{Gurtovenko89}:
\[
  \begin{array}{cc}
    V= {\rm const} = 1.08~{\rm~km/s}, &  H< 130~{\rm km}; \\
    V ={\rm const} = 0.55~{\rm~km/s}, &  H> 420{\rm~km} ; \\
    V= 0.946 + H(1.88 \cdot 10^{-3}- 6.75 \cdot 10^{-6} H  )~{\rm km/s}, & 130~{\rm km} < H < 420~{\rm km}.

\end{array}
\]

%___________________________________ Table 2
%
{\footnotesize
 \begin{table}[t] \centering
 \parbox[b]{14cm}{
\caption{ Response of line central depths ($\Delta R_i = R -R_i$) to temperature
disturbance. $R$, $R_i$ are line depths calculated for undisturbed and disturbed
models, $i$ is disturbance magnitude in percent.
 \label{T:1}}
\vspace{0.3cm}}
 \footnotesize
\begin{tabular}{ccccccccccc}
 \hline
 $\lambda $, nm & $EP$, eV & $R$ &$\Delta R_1$& $\Delta R_2$ &$\Delta R_4$&
$\Delta R_6$&$\Delta R_8$&$\Delta R_{10}$&$\Delta R_{15}$&$\Delta R_{20}$\\
 \hline

434.723& 0.00& 0.710& 0.050& 0.103& 0.208& 0.303& 0.382& 0.446& 0.552& 0.612 \\

Fe I   &     &      & 0.046& 0.092& 0.183& 0.275& 0.366& 0.458& 0.687& 0.916 \\

448.974& 0.12& 0.908& 0.009& 0.019& 0.041& 0.066& 0.094& 0.124& 0.208& 0.301 \\

Fe I   &     &      & 0.009& 0.017& 0.035& 0.052& 0.070& 0.087& 0.130& 0.174 \\

512.767& 0.05& 0.280& 0.038& 0.072& 0.126& 0.165& 0.194& 0.214& 0.244& 0.259 \\

Fe I   &     &      & 0.038& 0.076& 0.153& 0.229& 0.306& 0.382& 0.573& 0.764 \\

525.020& 0.12& 0.786& 0.020& 0.044& 0.100& 0.167& 0.241& 0.310& 0.465& 0.572 \\

Fe I   &     &      & 0.018& 0.037& 0.074& 0.110& 0.147& 0.184& 0.276& 0.368 \\

547.316& 4.19& 0.279& 0.018& 0.036& 0.067& 0.093& 0.115& 0.134& 0.169& 0.194 \\

Fe I   &     &      & 0.017& 0.034& 0.068& 0.101& 0.135& 0.169& 0.253& 0.338 \\

577.845& 2.59& 0.300& 0.027& 0.051& 0.094& 0.130& 0.158& 0.180& 0.219& 0.244 \\

Fe I   &     &      & 0.026& 0.051& 0.103& 0.154& 0.206& 0.257& 0.385& 0.514 \\

590.567& 4.65& 0.617& 0.018& 0.036& 0.074& 0.111& 0.147& 0.180& 0.253& 0.312 \\

Fe I   &     &      & 0.016& 0.033& 0.066& 0.098& 0.131& 0.164& 0.246& 0.328 \\

595.669& 0.86& 0.638& 0.034& 0.071& 0.148& 0.223& 0.290& 0.348& 0.452& 0.517 \\

Fe I   &     &      & 0.031& 0.062& 0.125& 0.187& 0.250& 0.312& 0.468& 0.624 \\

627.022& 2.86& 0.593& 0.024& 0.049& 0.102& 0.153& 0.201& 0.245& 0.332& 0.397 \\

Fe I   &     &      & 0.023& 0.045& 0.091& 0.136& 0.182& 0.227& 0.340& 0.454 \\
   \hline
 ~~~~~~~~~$\Delta_T$&(weak)    &   &0.001& 0.002& 0.012& 0.032& 0.060& 0.093& 0.193& 0.306 \\
 ~~~~~~~~~$\Delta_T$&(moderate)&   &0.002& 0.005& 0.014& 0.022& 0.025& 0.023& 0.010& 0.060 \\
 ~~~~~~~~~$\Delta_T$&(strong)  &   &0.001& 0.006& 0.019& 0.033& 0.045&0.058&0.134&0.212\\
    \hline
 \end{tabular}
 \end{table}}
 \noindent
 %%%%%%%%%%%%%%%%%%%%%%%%%%%%%%%%%%%%%%%%%%%%%%%%%%%%%

%___________________________________ Table 3
%

{\footnotesize
 \begin{table} \centering
 \parbox[b]{14cm}{
\caption{Response of Line Central Depths to Gas Pressure Disturbance.
 \label{T:3}}
\vspace{0.3cm}}
 \footnotesize
\begin{tabular}{ccccccccc}

 \hline
 $\lambda $, nm & $EP$, eV & $R$ &$\Delta R_1$& $\Delta R_5$ &$\Delta R_{10}$&
$\Delta R_{15}$&$\Delta R_{20}$&$\Delta R_{30}$\\
 \hline

463.407& 4.05& 0.720& 0.002& 0.008& 0.016& 0.023& 0.031& 0.047     \\
Cr II& & & 0.002& 0.012& 0.023& 0.035& 0.046& 0.070               \\
505.214& 7.68& 0.310& 0.003& 0.011& 0.022& 0.032& 0.042& 0.061    \\
C I& & & 0.003& 0.014& 0.028& 0.042& 0.056& 0.084                 \\
525.020& 0.12& 0.786& 0.001& 0.004& 0.008& 0.012& 0.017& 0.026    \\
Fe I& & & 0.001& 0.004& 0.008& 0.012& 0.017& 0.026                \\
590.567& 4.65& 0.617& 0.001& 0.005& 0.009& 0.014& 0.020& 0.030    \\
Fe I& & & 0.002& 0.009& 0.018& 0.026& 0.035& 0.053                \\
595.669& 0.86& 0.638& 0.000& 0.001& 0.000& 0.001& 0.003& 0.009    \\
Fe I& & & 0.001& 0.006& 0.011& 0.017& 0.022& 0.034                \\
624.756& 3.89& 0.565& 0.002& 0.010& 0.020& 0.029& 0.038& 0.057    \\
Fe II& & & 0.003& 0.014& 0.028& 0.041& 0.055& 0.083               \\
634.709& 8.09& 0.407& 0.003& 0.014& 0.027& 0.040& 0.053& 0.077    \\
Si II& & & 0.003& 0.016& 0.032& 0.047& 0.063& 0.095               \\
777.196& 9.14& 0.363& 0.003& 0.012& 0.024& 0.035& 0.046& 0.067    \\
O I& & & 0.003& 0.014& 0.027& 0.041& 0.055& 0.082                 \\
789.637& 10.00& 0.171& 0.002& 0.009& 0.019& 0.027& 0.036& 0.051   \\
Mg II& & & 0.002& 0.010& 0.020& 0.030& 0.040& 0.060               \\

   \hline
 ~~~~~~~~~$\Delta_P$&(weak)     &   & 0.000& 0.002& 0.005& 0.006& 0.009& 0.016 \\
 ~~~~~~~~~$\Delta_P$&(moderate) &   &0.001& 0.004& 0.008& 0.012& 0.015& 0.023\\
 ~~~~~~~~~$\Delta_P$&(strong)   &   &0.000& 0.003& 0.006& 0.009& 0.012&0.017\\
    \hline
 \end{tabular}
 \end{table}}
 \noindent
 %%%%%%%%%%%%%%%%%%%%%%%%%%%%%%%%%%%%%%%%%%%%%%%%%%%%%

%___________________________________ Table 4
%

{\footnotesize
 \begin{table} \centering
 \parbox[b]{15.5cm}{
\caption{Response of central depths and equivalent widths of lines to
microturbulent velocity disturbance ($\Delta W_i = W_i-W$, where $W$, $W_i$ are
equivalent widths calculated for undisturbed and disturbed models, $i$ is
disturbance magnitude in percent)
\label{T:4}} \vspace{0.3cm}}
 \footnotesize
\begin{tabular}{ccccccccccc}

 \hline
 $\lambda $, nm & $EP$, eV & $R$ &$W$,\,pm& $\Delta R_{30}$& $\Delta R_{50}$ &$\Delta R_{100}$&
$\Delta W_{10}$&$\Delta W_{30}$&$\Delta W_{50}$&$\Delta W_{100}$\\
 \hline
434.723& 0.00& 0.710& 3.753& 0.028& 0.049& 0.100& 0.065& 0.208& 0.336&0.639       \\
Fe~I& & & & 0.027& 0.045& 0.090& 0.066& 0.197& 0.329& 0.658                      \\
438.925& 0.05& 0.890& 7.429& 0.002& 0.004& 0.008& 0.244& 0.726& 1.236&2.592      \\
Fe~I& & & & 0.002& 0.003& 0.006& 0.236& 0.708& 1.180& 2.360                      \\
448.974& 0.12& 0.908& 9.579& 0.001& 0.002& 0.004& 0.276& 0.872& 1.512&3.241      \\
Fe~I& & & & 0.001& 0.002& 0.004& 0.324& 0.972& 1.620& 3.240                      \\
508.334& 0.96& 0.865& 10.768& 0.001& 0.003& 0.006& 0.287& 0.906& 1.577&3.405     \\
Fe~I& & & & 0.001& 0.002& 0.004& 0.364& 1.092& 1.820& 3.640                      \\
512.767& 0.05& 0.280& 1.449& 0.024& 0.040& 0.074& 0.006& 0.016& 0.039& 0.069     \\
Fe~I& & & & 0.025& 0.042& 0.084& 0.006& 0.019& 0.032& 0.063                      \\
525.020& 0.12& 0.786& 6.430& 0.006& 0.010& 0.024& 0.192& 0.562& 0.959& 1.935     \\
Fe~I& & & & 0.005& 0.009& 0.018& 0.178& 0.534& 0.890& 1.780                      \\
547.316& 4.19& 0.279& 1.743& 0.022& 0.036& 0.068& 0.009& 0.044& 0.061& 0.110     \\
Fe~I& & & & 0.024& 0.040& 0.080& 0.008& 0.023& 0.039& 0.077                      \\
577.845& 2.59& 0.300& 1.875& 0.024& 0.042& 0.074& 0.011& 0.047& 0.067& 0.120     \\
Fe~I& & & & 0.025& 0.042& 0.084& 0.011& 0.033& 0.053& 0.112                      \\
590.567& 4.65& 0.617& 6.084& 0.013& 0.022& 0.043& 0.170& 0.408& 0.653& 1.289     \\
Fe~I& & & & 0.013& 0.022& 0.040& 0.146& 0.438& 0.730& 1.458                      \\

   \hline
 ~~~~~~~~~$\Delta_V$&(weak)    &  & &0.000& 0.002& 0.011& 0.000& 0.013& 0.013&0.016\\
 ~~~~~~~~~$\Delta_V$&(moderate)& & &0.000& 0.004& 0.010& 0.001& 0.011&0.007& 0.019\\
 ~~~~~~~~~$\Delta_V$&(strong)  & & &0.000& 0.019& 0.002& 0.034&0.083& 0.119&0.130\\
    \hline
 \end{tabular}
 \end{table}}
 \noindent
 %%%%%%%%%%%%%%%%%%%%%%%%%%%%%%%%%%%%%%%%%%%%%%%%%%%%%

To construct the disturbed models, we assumed that one parameter only, $T$ for
instance, varied by $\Delta T_i =(i/100)T$ due to the disturbance ($i$ is the
disturbance magnitude in percent). Then the new temperature value in the disturbed
model becomes $T_i(H) = T(H) + \Delta T_i(H)$. Other model parameters are
calculated from $H,~T_i,~P,~V$. Tables 2--4 give the results of calculations for
the variations in the depression at the line center ($\Delta R_i$) arising due to a
disturbance of the model parameters $T, P$, and $V$. The first row for each line
gives the values of $\Delta R_i = R- R_i$ obtained from direct calculations of
$R_i$ for a disturbed model and $R$ for the undisturbed one. The second row gives
the values of $\Delta R$ calculated with the sensitivity parameters from (10). The
lines are selected in such a way that their parameters $EP,~R,~P_R,~\beta$ be as
diverse as possible. The last three rows in the tables give mean absolute values
for the differences $\Delta_{\beta} = | \Delta R_i- \delta R |$ separately for
weak, moderate, and strong lines.

The analysis of the results reveals that the magnitudes of response calculated
directly ($\Delta R_i$) begin to deviate for certain values of $ \delta\beta
/\beta$ from the linear dependence of $ \delta R$ on $ \delta\beta/\beta$ obtained
with (10). The quantity $\Delta_{\beta}$ just characterizes this deviation. When
the initial temperature is disturbed, the quantity $\Delta_T$ increases with the
magnitude of disturbance much faster than for the disturbances of the gas pressure
or microturbulent velocity. The quantity $\Delta_T$ depends on the excitation
potential and line strength, reaching its lowest value for moderate lines with
large $EP$. It is evident from Table 2 that the sensitivity parameters may be used
for the diagnostics of the solar photosphere when the observed variation $\delta R$
is no more than $R/2$ and the disturbance magnitude is no more than 8\%
($\approx400$~K). There must be no disturbance inversion in the line formation
region in this case.

We have somewhat another picture for the gas pressure disturbance. The sensitivity
of lines to changes in $P$ is much smaller, and the quantities $\Delta_P$ are
small, and therefore the sensitivity parameters may be used with confidence for
estimating $\delta R$ or $\delta P$ when the pressure disturbance is 30\% or more.
The quantities $\Delta_V$ for central depths are insignificant at the
microturbulent velocity disturbances up to 50\%. For equivalent widths, $\Delta_V$
becomes more significant, being greater for stronger lines. The deviations for
strong lines attain 0.13 pm when $V$ increases by 100\%.

If $\delta\beta/\beta$ depends on height or the disturbance is localized in regions
much smaller than the region of formation of the absorption line, the inverse
problem cannot be solved, i.e., the quantity $\delta\beta/\beta$ cannot be
determined from the observed $\delta R/R$ with (11). Formulae (1), (2) should be
used in this case.

Now we dwell on the determination of the height where the effective response to the
line depression variations occurs, resulting from an atmospheric parameter
disturbance. This is easy to do with the response functions. One has to find the
height at which the center of gravity of the integrand in (1) lies. We have done
such calculations for the lines given in Tables 2--4, with the results shown in
Table 5, where $h_{R,\beta}$ and $h_{W,\beta}$ are the heights of the effective
response to variations in $T$, $P$, and $V$ of the line center depression and
equivalent width calculated from the depression response functions; $h_R$ and $h_W$
are the heights of effective formation of depression at the line center and for the
line as a whole, they are calculated from the depression contribution functions.
The heights $h_R$ and $h_{R,\beta}$, as well as $h_W$ and $h_{W,\beta}$ are
obviously different.

%___________________________________ Table 5
%

{\footnotesize
 \begin{table} \centering
 \parbox[b]{15.5cm}{
\caption{Heights of formation of effective response of lines.
 \label{T:5}}
\vspace{0.3cm}}
 \footnotesize
\begin{tabular}{lrrrrrrrr}

 \hline
 $\lambda $,\,nm & $h_R$,\,eV & $h_{R,T}$ &$h_{R,V}$ &$h_{R,P}$& $h_W$ &$h_{W,T}$&
$h_{W,V}$&$h_{W,P}$\\
 \hline
434.723~  Fe I& 271& 240& 205& 328& 232& 207& 179& 228         \\
438.925~  Fe I& 454& 467& 484& 406& 327& 227& 165& 258        \\
448.974~  Fe I& 560& 575& 616& 513& 371& 223& 156& 249        \\
463.407~  Fe II& 207& 178& 202& 172& 146& 122& 83& 106        \\
505.214~  C I& 59& 99& 66& 53& 41& 73& 35& 39                \\
512.767~  Fe I& 225& 222& 193& 268& 208& 208& 186& 247        \\
525.020~  Fe I& 373& 372& 342& 323& 289& 233& 179& 256        \\
547.316~  Fe I& 147& 143& 133& 96& 130& 128& 121& 92          \\
577.845~  Fe I& 182& 178& 160& 158& 165& 163& 148& 155        \\
590.567~  Fe I& 230& 216& 214& 186& 169& 142& 118& 132        \\
595.669~  Fe I& 289& 269& 234& 235& 240& 213& 178& 205        \\
624.756~  Fe II& 214& 222& 212& 186& 158& 158& 101& 129       \\
627.022~  Fe I& 253& 239& 219& 203& 204& 181& 150& 162        \\
634.709~  Si I& 109& 164& 121& 108& 76& 111& 45& 76           \\
777.196~  O I& 125& 183& 138& 121& 86& 113& 49& 86           \\
789.637~  Mg II& 57& 98& 63& 64& 40& 61& 34& 52                \\
    \hline
 \end{tabular}
 \end{table}}
 \noindent
 %%%%%%%%%%%%%%%%%%%%%%%%%%%%%%%%%%%%%%%%%%%%%%%%%%%%%

To find lines which are the most sensitive to one atmospheric parameter and at the
same time have low sensitivity to other parameters, we calculated and compared the
relative variations of line parameters, $\Delta R/R$, $\Delta W/W$, $\Delta R_ {
\Delta \lambda_{R{/}2} }{/} (R{/}2)$, rather than sensitivity indicators, for
actually observed fluctuations in the temperature (2\%), gas pressure (30\%), and
microturbulent velocity (50\%). This way proved to be appropriate for finding the
necessary lines. Here we give as an example the wavelengths of typical highly
sensitive lines in respect to their relative sensitivity. The most responsive to
temperature are the lines 670.778 Li I, 657.279 Ca I, 623.936 Sc I, 546.049 Ti I,
482.745 V I, 511.248 Cr I, 543.253 Mn I, 512.768 Fe I, 662.502 Fe I, 673.952 Fe I,
612.026 Fe I, 635.384 Fe I, 671.032 Fe I, 680.186 Fe I, 685.164 Fe I, 401.109 Co I,
412.830 Y I; the most responsive to gas pressure are the lines 477.588 C I, 505.214
C I, 658.762 C I, 711.317 C I, 833.516 C I, 824.252 N I, 777.196 O I, 777.539 O I,
926.590 O I, 871.783 Mg I, 789.637 Mg II, 875.202 Si I, 634.709 Si II, 891.207 Ca
II, 989.070 Ca II, 463.407 Cr II, 463.531 Fe II, 624.756 Fe II; and to
microturbulent velocity: 408.295 Mn I, 438.925 Fe I, 448.974 Fe I, 508.334 Fe I,
512.735 Fe I, 514.174 Fe I, 522.553 Fe I, 550.678 Fe~I, 625.256 Fe I, 633.534 Fe
I, 450.828 Fe II, 402.090 Co I, 508.111 Ni I, 464.866 Ni I, 510.554 Cu I, 481.053
Zn I, 460.733 Sr I, 488.369 Y II, 420.898 Zr II, 455.404 Ba II, 649.691 Ba II.

 Analyzing the starting data of these lines and the calculated values
of $\Delta R/R$, $\Delta W/W$, $\Delta R_ { \Delta \lambda_{R{/}2} }{/} (R{/}2)$,
we can conclude that the most responsive to temperature are the lines with
excitation potentials from 0 to 2 eV, central depths up to 0.35, equivalent widths
to 3 pm. Relative variations of the equivalent widths of these lines are 25--30\%
when the temperature changes by 2\%, while they are 2--7\% when the gas pressure
changes by 30\% and microturbulent velocity by 50\%. Moderate-strong lines of light
atoms with $EP\geq 6$ eV are highly responsive to gas pressure. For them, $\Delta
W/W$ is as much as 25--48\% when $P$ changes by 30\% and is only 2--6\% when $T$
changes by 2\% and $V$ by 50\%. The equivalent widths and half-width depths of
strong lines of heavy atoms with $W \approx 8$--12 pm and $\Delta\lambda_{R/2}
\approx 4$--6 pm are very sensitive to the microturbulent velocity. For these
lines, $\Delta W/W$ amounts to 11--18\% and $\Delta (R/2){/}(R/2)$ runs as high as
45--70\% when $V$ changes by 50\%, and it is 2\% when $P$ changes by 30\% and $T$
by 2\%.

 %%==========================================================

\section{Conclusion}

As we could make sure, the sensitivity of every line depends on numerous atom and
line parameters as well as on the model atmosphere parameters. It is measured by a
diversity of sensitivity indicators with which it is possible to determine the
atmospheric parameter producing the strongest response of the line and to estimate
the magnitude of the response. It should be taken into consideration which
sensitivity, the absolute or relative one, is more suitable for the particular
problem. Our calculations reveal that the lines which form under the LTE conditions
in the solar atmosphere are the most responsive to temperature variations. Evidence
of this is found, for example, in the values of parameters of relative sensitivity
to temperature, gas pressure, and microturbulent velocity which attain 15, 1.5,
1.5, respectively (see Table 1). Even minor temperature fluctuations, of the order
of 1\%, can be observed in the most responsive lines. Fluctuations in the density
of matter or in the gas pressure as well as variations in the microturbulent
velocity change the line only slightly as compared to temperature. Using the lines
most responsive to the gas pressure, it is possible to detect only 10--15\%
fluctuations of the pressure in the line formation region. Variations in the
microturbulent velocity can be measured from variations in the half-widths and
equivalent widths of highly sensitive lines if $\Delta V{/}V$ is about 30\%.

Direct calculations of the response of line depression to variations in atmospheric
parameters for disturbed and undisturbed model atmospheres, confirm that the
sensitivity indicators can be used for the solar atmosphere diagnostics in those
cases when fluctuations in $T$, $P$, and $V$ do not exceed 8\%, 50\%, and 100\%,
respectively.

It is better to use the response functions themselves when fine spectral analysis
problems are solved. A detailed analysis of the shape of response functions can
give an additional valuable inftyormation. The results of our analysis show that
the sensitivity indicators proposed for measuring the Fraunhofer line response do
not lack a physical meaning. We think it reasonable to include the sensitivity
parameters in the initial data for spectral lines together with the central depth,
equivalent width, and other principal characteristics when the data bases such as
``The Fraunhofer Spectrum'' \cite{Gadun92} are compiled.

%========-------------------------------------------------

{\bf Acknowledgements.}
The author thanks V. N. Karpinskii for valuable remarks and
useful discussion.

%%%%%%%%%%%%%%%%%%%%%%%%%%%%%%%%%%%%%%%%%%%%%%%%%%%%%%%%%%%%

\newpage

\end{document}